\documentclass{article}

\usepackage{amsmath,amssymb,amsfonts,amsthm}
\usepackage{bm,bbm}
\usepackage{graphicx}
\usepackage{esint}
\usepackage{color}
\usepackage{hyperref}

\newcommand{\be}{\begin{equation}}
\newcommand{\ee}{\end{equation}}
\newcommand{\ba}{\begin{eqnarray}}
\newcommand{\ea}{\end{eqnarray}}
\def\bs{\begin{subequations}}
\def\es{\end{subequations}}
\def\a{\alpha}
\def\b{\beta}
\def\de{\delta}
\def\g{\gamma}
\def\la{\lambda}

\def\e{\epsilon}
\def\ve{\varepsilon}
\def\Om{\Omega}
\def\om{\omega}

\def\s{\sigma}
\def\vr{\varrho}

\def\N{\nabla}

\def\cC{{\cal C}}
\def\cD{{\cal D}}
\def\cE{{\cal E}}
\def\cF{{\cal F}}

\def\cK{{\cal K}}
\def\cL{{\cal L}}
\def\cM{{\cal M}}

\def\cO{{\cal O}}
\def\cP{{\cal P}}

\def\cS{{\cal S}}

\def\cV{{\cal V}}
\def\bE{\mathbbm{e}}

\def\ds{d_{\rm S}}
\def\dh{d_{\rm H}}
\def\dw{d_{\rm W}}

\def\p{\partial}
\def\bp{\bar{\partial}}

\newcommand{\Eq}[1]{(\ref{#1})}

\def\cob{\color{blue}}
\def\cor{\color{red}}
\newcommand{\oarX}[1]{\href{http://arxiv.org/abs/#1}{{\ttfamily\cor #1}}}
\newcommand{\arX}[1]{\href{http://arxiv.org/abs/#1}{{\ttfamily\cor arXiv:#1}}}
\newcommand{\doin}[5]{\href{http://dx.doi.org/#1}{\cob {\it #2} {\bf #3}, #4 (#5)}}
\newcommand{\doij}[5]{\href{http://dx.doi.org/#1}{\cob {\it #2} #3 (#5) #4}}
\newcommand{\ndoin}[5]{\href{#1}{\cob {\it #2} {\bf #3}, #4 (#5)}}
\newcommand{\tia}[1]{}

\def\lp{\ell_{\rm Pl}}
\def\rme{e}
\def\rmd{d}
\def\rmi{i}

\begin{document}

\title{Introduction to multifractional spacetimes}



\author{Gianluca Calcagni\\
\small Max Planck Institute for Gravitational Physics (Albert Einstein Institute),\\
\small Am M\"uhlenberg 1, D-14476 Golm, Germany\\
\small Instituto de Estructura de la Materia, IEM-CSIC, Serrano 121, 28006 Madrid, Spain\\
calcagni@iem.cfmac.csic.es}

\maketitle

\begin{abstract}
We informally review the construction of spacetime geometries with multifractal and, more generally, multiscale properties. Based on fractional calculus, these continuous spacetimes have their dimension changing with the scale; they display discrete symmetries in the ultraviolet and ordinary Poincar\'e symmetries in the infrared. Under certain reasonable assumptions, field theories (including gravity) on multifractional geometries are generally argued to be perturbatively renormalizable. We also sketch the relation with other field theories of quantum gravity based on the renormalization group.
\end{abstract}

\date{August 9, 2012}

\medskip

\doin{10.1063/1.4756961}{AIP Conf.\ Proc.}{1483}{31}{2012}\hfill \arX{1209.1110}



\section{Motivation}


\subsection{Some numerology}

2 is a recurrent number in quantum gravity. In the first attempts to formulate a perturbative theory of gravity, it was found that the latter was renormalizable near $D=2$ dimensions (e.g., \cite{calcagni:2e}); the hope, later unfulfilled, was that gravity would be renormalizable in $D=2+\ve$ dimension also in the limit $\ve\to 2$. String theory is defined on a two-dimensional manifold, the worldsheet, where powerful conformal techniques are available and interactions (and, hence, ultraviolet divergences) are smeared over spacetime. In other scenarios, correlation functions change behavior as across a phase transition and the dimension of spacetime varies with the scale, a phenomenon known as dimensional reduction or dimensional flow \cite{calcagni:df}. This feature is, in a broad sense, universal in quantum gravity and it is related to the ultraviolet (UV) finiteness of the theories. Examples are noncommutative geometry (both at the fundamental and effective level) \cite{calcagni:ncg,calcagni:ACOS}, loop quantum gravity and spin-foams (where the spectral dimension is somewhat close to $\ds\sim 2$ in the UV) \cite{calcagni:sf}, asymptotic safety (where $\ds=2$ in the UV and, in certain models, there is also an intermediate regime with $\ds\sim 4/3$) \cite{calcagni:as,calcagni:asf}, causal dynamical triangulations \cite{calcagni:cdt}, Ho\v{r}ava--Lifshitz gravity (where $\ds=2$ in the UV) \cite{calcagni:hl}, and other approaches \cite{calcagni:qs}.

One should also mention a curious remark based on elementary dimensional arguments \cite{calcagni:Bar83}. Including Planck's constant $\hbar$, Newton's constant $G$, the speed of light $c$, and the electron charge $e$,  one can construct a dimensionless constant in a spacetime of topological  dimension $D$ as $C=\lp^{2(3-D)} e^{D-2} G^{{D}/{2}-1} c^{2(2-D)}$, where $\lp:=\sqrt{\hbar G/c^3}$ is Planck's length. The same formula holds if one replaces the topological dimension with the Hausdorff dimension $\dh$. Notably, in $\dh=2$ the fundamental constant coincides with (the square of) the Planck length, $C=\lp^2$, while all the other couplings disappear. This observation is made all the more mysterious by the presence of the electric charge, suggesting that, if the constant $C$ were related to a concrete quantum gravity theory, the latter should automatically involve also matter.

A series of questions may come to the mind of the reader at this point. Why do two dimensions play such a role? Why do we have 4 dimensions in the infrared (IR)? If these questions were not meaningful by their own separately, one might attempt to answer the following: \emph{Why 2 and 4 dimensions}? More generally, how to control the details of dimensional flow? Multifractional spacetimes will be able to address the last two queries.


\subsection{From fractal to multifractional spacetimes}

Traditional perturbative field theory fails to quantize gravity consistently. Other, more advanced frameworks which are nonperturbative and are based on a discrete pre-geometric structure (such as loop quantum gravity, spin-foams and simplicial gravity) are more successful but then fail to fully recover a large-scale, continuum, classical picture. At the same time, these theories display dimensional flow in various incarnations. From these sparse observations, one might try a reactionary step back to a continuum perturbative field theory, implement dimensional flow therein, and see if renormalization of gravity comes as a byproduct. The recipe to achieve this is the following:
\begin{enumerate}
\item[(i)] The formalism should describe dimensional flow and other features of quantum-gravity theories with tools borrowed from other branches of physics and mathematics. 
\item[(ii)] Dimensional flow can be realized at the structural level (rather than as an indirect property). 
\item[(iii)] It should be defined on a continuous geometric structure.
\item[(iv)] Gravity should be (power-counting) renormalizable.
\item[(v)] The system should be invariant under some symmetry group and Lorentz invariance should be recovered at large scales.
\end{enumerate}
We comment each element separately. (i) Multiscale phenomena and geometries are best studied in the theory of complex systems and in multifractal geometry \cite{calcagni:mss}. In this context, one has several definitions of dimension. The spectral dimension  $\ds$ is a somewhat indirect geometric indicator, because it is found via diffusion of a pointwise source probing local geometry. On the other hand, the Hausdorff dimension $\dh$ is an immediate characteristic of the measure in position space and, hence, of the geometric construction. Looking at dimensional flow also at the level of the Hausdorff dimension, one realizes point (ii). However, attempting to construct a field theory directly on a fractal or multifractal turns out to be very difficult because of the extreme disconnectedness of the ``medium'' (see the discussion and references in \cite{calcagni:frc2} for early attempts), although a field formalism can be built on graphs representing discrete geometries \cite{calcagni:COT1}. Calculus on the discrete analogue of differential manifolds is still in a phase of development, so one could resort to a continuum geometry (iii). Then, the proposal acquires the double character of being a \emph{fundamental} theory or an \emph{effective} description of certain regimes of other, discrete models. In the first case, one must check several properties, including renormalizability at the perturbative level of otherwise pathological field theories (iv). Finally, one should make sure that violation of ordinary Poincar\'e symmetries in the UV is not enhanced at the quantum level \cite{calcagni:col}, as it happens for Lifshitz-type models (and, most probably, also for Ho\v{r}ava--Lifshitz gravity) \cite{calcagni:IRS}.

Concretely, a simple implementation of dimensional flow (multifractal geometry) is a \emph{change of measure} of position space:
\be 
d^Dx\to d\vr(x)\,,
\ee
where $D$ is the topological dimension of spacetime (we do not set $D=4$ for the moment) and $\vr$ is a generic Lebesgue--Stieltjes measure. As we said, general fractal measures may become rapidly intractable, but this may be true also for arbitrary continuous measures. Even for a Lebesgue measure $d\vr(x)=d^Dx\,v(x)$, it soon becomes clear that a number of conceptual and quantitative problems cannot be tackled if the function $v(x)$ is not factorizable in the coordinates \cite{calcagni:fra,calcagni:frc3,calcagni:frc5}.

Some mathematical results in one dimension bridge the gap between fractals and continuous measure \cite{calcagni:ff}. Under certain approximations \cite{calcagni:frc2,calcagni:frc1}, calculus on fractals can be replaced by continuous \emph{fractional} calculus. Therefore, it is natural to consider fractional integrals over a space with fractional dimension as models of geometries with fractal properties. Extending the picture to many dimensions, Lorentzian signature and scale-dependent geometry, one obtains a model of \emph{multifractional spacetimes} where dimensional flow is only one among many interesting properties.

Multifractional spacetimes have been introduced in \cite{calcagni:fra4}. Fractional Euclidean and Minkowski spaces are constructed in \cite{calcagni:frc1} and \cite{calcagni:frc2}, respectively. Momentum space and the fractional analogue of Fourier transform are defined in \cite{calcagni:frc3}. The generalization to multiscale spaces and diffusion equation are discussed in \cite{calcagni:frc2,calcagni:fra6}. Power-counting renormalizability of field theories on these spaces has begun to be studied in \cite{calcagni:frc2}, although the details of the quantum theory are still work in progress. Nevertheless, various applications have been already formalized, such as in noncommutative geometry \cite{calcagni:ACOS}, quantum mechanics \cite{calcagni:frc5}, asymptotic safety \cite{calcagni:asf}, and, in a toy model closer to \cite{calcagni:fra}, the Standard Model \cite{calcagni:HX}.

Several added \emph{boni} arise from this construction. Apart from the connection with noncommutative spacetimes, a physical clarification of $\kappa$-Minkowski spacetime \cite{calcagni:ACOS}, and some insights into renormalization-based approaches to quantum gravity \cite{calcagni:asf}, we get a natural discrete-to-continuum transition of geometry and the emergence of a scale hierarchy \cite{calcagni:frc2}. The spectral theory and the Fourier transform on fractals, a still underdeveloped branch of mathematics, receives a full treatment in the more limited context of fractional geometries \cite{calcagni:frc3,calcagni:fra6}. Also, one is able to clarify the relationship between multiscale geometries, stochastic processes, and analytic profiles for the spectral dimension $\ds$ in quantum gravity \cite{calcagni:asf}. 

In this paper, we will review the formalism and some of these developments. The goal is to elicit the reader's interest in a theory at its early stages but which can be advanced in a rather robust and direct way thanks to (adaptations of) familiar techniques in continuum calculus. 

Before beginning, it may be instructive to consider a set of questions and caveats which may spontaneously arise at this point.
\begin{itemize}
\item\emph{Q1: In what sense do these models live on a ``(multi)fractal?''}

\emph{A1:} Dimensional flow is smooth, thus implying transitions through states with noninteger $\dh$ and/or $\ds$.
\item\emph{Q2: Are there fractals with integer $\dh$ and/or $\ds$?}

\emph{A2:} Yes. Deterministic examples with $\dh=2$ are plane-filling curves (dragon, Moore, Peano, Sierpi\'nski curve), the Mandelbrot set and its boundary, the Sierpi\'nski tetrahedron, Pythagoras tree, and space-filling curves (Moore, Hilbert, Lebesgue curve). In particular, diamond fractals can have $\dh=\ds=2$. Other instances can be found among random fractals (such as the trail and graph of Brownian motion) or in various natural fractals.
\item\emph{Q3: Is it really necessary to imagine this model on a ``(multi)fractal?'' Are all spaces with anomalous dimension fractals?}

\emph{A3:} No and No. Realization of dimensional flow is what matters. In fact, there exist geometric configurations which cannot be classified as fractal. If $\ds>\dh$, they are associated with jump processes \cite{calcagni:frc1,calcagni:fra6}.
\item\emph{Q4: Does $\dh=2$ and/or $\ds=2$ in the UV guarantee renormalizability?}

\emph{A4:} No. A detailed renormalization-group (RG) analysis is required. Yet, power-counting renormalizability is already a positive indication of the UV finiteness of multifractional field theories \cite{calcagni:frc2}.

\item\emph{Q5: What is the meaning of ``dimension?''}

\emph{A5:} Aside from various operational definitions of dimension (topological, Hausdorff, spectral, walk, box, and so on), it is not obvious how, in an anomalous geometry, a varying dimension is related to physics (and physical degrees of freedom). In Sec.\ \ref{calcagni:mqg}, we establish a formal duality between the multifractional description and gravitational field theories based on renormalization techniques (e.g., asymptotic safety and Ho\v{r}ava--Lifshitz gravity). As a byproduct, we will show that \emph{the very concept of dimension is, in fact, the notion of adapted rod}. By ``adapted rod'' one means, more precisely, scale-dependent physical momentum.
\end{itemize}


\section{Fractional Euclidean space}

We begin with the simplest fractional space, the counterpart of Euclidean space. Fractional Euclidean space has no time direction and its dimension is fixed at all scales.


\subsection{Definition and measure}

Fractional Euclidean space of real order $\a$ is specified by the set of data
\be\label{calcagni:def}
\cE_\a^D = ({\mathbb{R}^D},\,{\vr_\a},\,{\rm Calc}^\a,\,{\| \cdot \|},\,{\cK})\,.
\ee
The first entry is the embedding space, in this case ordinary $D$-dimensional Euclidean space $\mathbb{R}^D$. $\vr_{\a}$ is the measure in position space (appearing in the action) and it is associated with a specification ${\rm Calc}^\a$ of the differential structure and calculus. The space is endowed with a natural norm $\| \cdot \|$ and a Laplacian operator $\cK$.

Without loss of generality, we consider a ``bilateral'' (i.e., with support also on negative-valued coordinates) and ``isotropic'' (same order $\a$ for all directions) measure
\be\label{calcagni:mea}
\rmd\vr_\a(x)=\rmd^D x\,v_\a(x)=\rmd^D x\, \prod_\mu \frac{|x^\mu|^{\a-1}}{\Gamma(\a)}\,,
\ee
where $\Gamma$ is the gamma function and $\mu=1,\dots,D$. We sometimes call $\a$ fractional charge. Ordinary integration is thus replaced by
\be\label{calcagni:fri}
\int_{-\infty}^{+\infty}\rmd^D x \to {\int_{-\infty}^{+\infty}\rmd\vr_\a(x)}\,.
\ee
Apparently, unilateral measures ($x^\mu\geq 0$) seem not to be fit for quantum mechanics and quantum field theory \cite{calcagni:frc5}.

The distribution $\vr_\a(x^\mu)=q^\mu:={|x^\mu|^\a}/{\Gamma(\a+1)}$ defines a set of ``geometric'' coordinates such that the integration measure formally reduces to the usual one, $\rmd\vr_\a=\rmd^D q$. The choice of $\{x\}$ or $\{q\}$ as the coordinates associated with rod measurements strongly relies on the form of the Laplacian and, above all, on the identification of physical momentum (see \cite{calcagni:asf} and Sec.\ \ref{calcagni:mqg}).

The measure \Eq{calcagni:mea} obeys the scaling property
\be\label{calcagni:scla}
\vr_\a(\la x)=\la^{D\a} \vr_\a(x)\,.
\ee
Anomalous scaling is thus {natural} in fractional and Lebesgue--Stieltjes integrals.


\subsection{Calculus}

Fractional manifolds are continua which are not differentiable with respect to ordinary calculus. The latter is replaced by fractional calculus, as old as the former and first developed by the same mathematicians (in particular, Leibniz, Riemann, and Liouville). Care must be taken to represent fractional operators and define functional calculus, but once this is properly done the final product is completely self-consistent \cite{calcagni:KST}. Applications include dissipative mechanics, chaos and percolation theory, anomalous transport systems \cite{calcagni:mss}, statistics and long-memory processes such as weather and stochastic financial models, and system modeling and control in engineering (see references in \cite{calcagni:frc1}). To the best of our knowledge, fractional spaces are the first systematic application of fractional calculus to quantum field theory and quantum gravity.

Examples of fractional operators are the left Caputo derivative
\be
(\p^\a f)(x) := \frac{1}{\Gamma(1-\a)}\int_{0}^{x} \frac{\rmd x'}{(x-x')^\a}\p_{x'}f(x')\,,\qquad 0<\a\leq 1\,,
\ee
and the Weyl derivative
\be
({}_\infty\bp^\a f)(x) :=-\frac{1}{\Gamma(1-\a)}\int_{x}^{+\infty}\frac{\rmd x'}{(x'-x)^\a}\p_{x'}f(x')\,,\qquad 0<\a\leq 1\,.
\ee
They generalize integer derivatives in a nonlocal fashion dependent on the integration domain. Right Caputo and ``left Weyl'' (called Liouville) derivatives also exist. A property preserved by Caputo, Weyl and Liouville derivatives is that their action on a constant gives zero. This may not be true for other derivatives such as Riemann--Liouville.

\begin{figure}
\includegraphics[width=9.6cm]{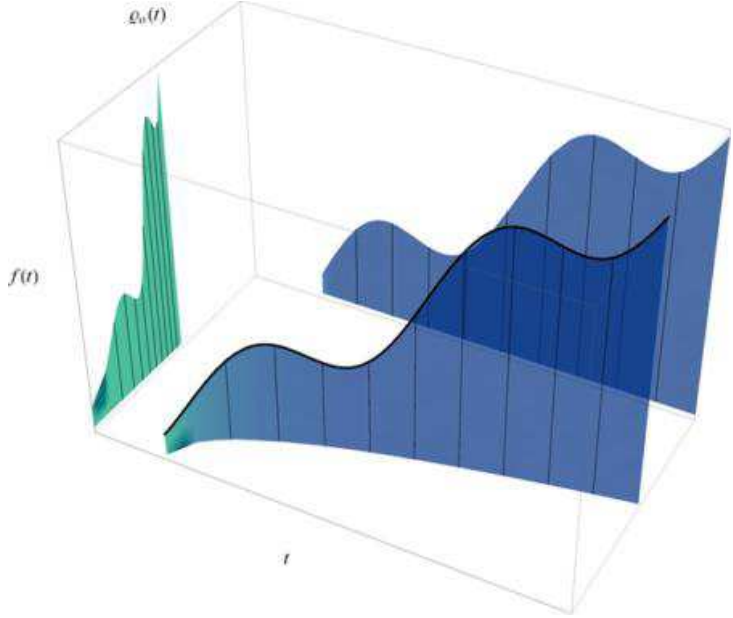}
\caption{Geometric interpretation of fractional integrals as ``shadows'' of a ``fence.'' In the figure, $\a=1/2$ \cite{calcagni:frc1}.}
\label{calcagni:fig1}
\end{figure}

The integral \Eq{calcagni:fri} can be recognized as a bilateral modification of the Weyl fractional integral. In one dimension, the latter is
\be
({}_\infty \bar I^\a f)(x_0) :=\frac{1}{\Gamma(\a)}\int_{x_0}^{+\infty}\rmd t\,(x-x_0)^{\a-1} f(x)\,.
\ee
Fractional integrals admit a neat geometric interpretation \cite{calcagni:BuPo}. Consider a function $f(t)$ and the left time integral
\be\label{calcagni:It}
(I^\a f)(t_1) =\int_{t_0}^{t_1}\rmd t\,\frac{(t_1-t)^{\a-1}}{\Gamma(\a)} f(t)=:\int_{t_0}^{t_1}\rmd t\,v_\a(t_1-t) f(t) =\int_{t_0}^{t_1} \rmd\vr_\a(t)\,f(t)\,.
\ee
The geometric meaning of the left fractional integral \Eq{calcagni:It} with $\a\neq 1$ fixed is shown in Fig.\ \ref{calcagni:fig1}. The continuous curve in the box is given parametrically by the set of points $\cC=\{(t,\vr_\a(t),f(t))\}$, where $f$ is some smooth function. Projection of $\cC$ onto the $t$-$f$ plane ($\vr_\a={\rm const}$) gives $f(t)$, while projection onto the $t$-$\vr_\a$  plane ($f={\rm const}$) yields $\vr_\a(t)$. Now, build a vertical ``fence'' under the curve $\cC$, and project it onto both planes. On the $t$-$f$ plane, the shadow of the fence is the ordinary integral, $(I^1 f)(t_1) =\int_{t_0}^{t_1} \rmd t\, f(t)$. In classical mechanics, it corresponds to a full-memory process, $\a=1$. On the $\vr_\a$-$f$ plane, the shadow corresponds to the fractional integral \Eq{calcagni:It}, the area under the projection of $\cC$ on such plane. The limit $\a\sim 0$ corresponds to a Markov (no-memory) process. $\a$ is then interpreted as the fraction of states preserved at a given time $t$. If $f$ was the velocity of a particle, the distance travelled in an ordinary and a fractional geometry would be different.


\subsection{Norm}

\begin{figure}
\includegraphics[width=9.6cm]{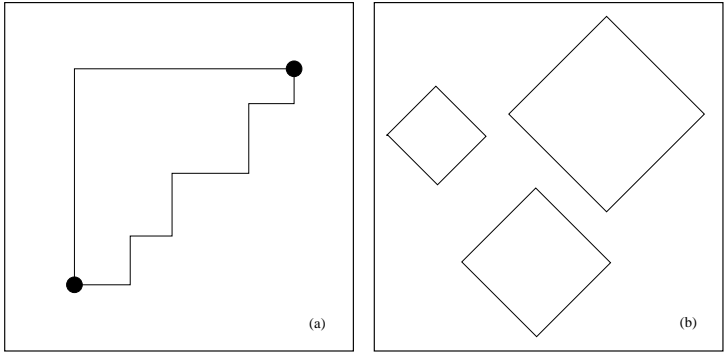}
\caption{1-norm and taxicab geometry in two dimensions. (a) Left panel: the shortest path between two points is not unique. (b) Right panel: circles of radius $R$ are diamonds with edges at $45^\circ$ with respect to the coordinate axes, $|x|+|y|=R$ \cite{calcagni:frc1}.}
\label{calcagni:fig2}
\end{figure}

From the perspective of fractional differential forms \cite{calcagni:fdf}, the $2\a$-norm is the natural distance \cite{calcagni:frc1}:
\be\label{calcagni:norm}
\Delta_\a(x,y) := \left[\sum_{\mu=1}^D\delta_{\mu\nu}(|x^\mu-y^\mu||x^\nu-y^\nu|)^{\a}\right]^{\frac{1}{2\a}}.
\ee
This is a norm only if $\a\geq 1/2$, i.e., when the triangle inequality holds. Therefore, we can {restrict} $\a$ to lie in the range
\be\label{calcagni:range}
\frac12\leq\a\leq1\,.
\ee
Notice that varying $\a$ in \Eq{calcagni:norm} does \emph{not} produce a topologically equivalent norm, since also the geometric texture would change. The norm defines the fractional $D$-ball of radius $R$, the locus of points no further than $R$ from a center. Fractional balls are obviously not rotation invariant due to the nontrivial measure. For instance, in the limiting case $\a=1/2$, $D=2$, one has the so-called taxicab or Manhattan geometry, where circles are diamonds and the shortest distance between two points is not unique (Fig.\ \ref{calcagni:fig2}).


\subsection{Laplacian}

It is possible to construct a self-adjoint fractional Laplacian of order $2\g$ \cite{calcagni:fra6}:
\be\label{calcagni:K}
\cK_{\g,\a} := \sum_\mu \cK_{\g,\a}(x^\mu)=-\frac{1}{\sqrt{v_\a(x)}}\,\frac{\sum_\mu({}_\infty\p^{2\g}_\mu+{}_\infty\bp^{2\g}_\mu)}{2\cos(\pi\g)}\left[\sqrt{v_\a(x)}\,\cdot\right],
\ee
where $0<\g\leq m$ and we made use of Liouville and Weyl fractional derivatives of higher order: for each direction, $({}_\infty\p^{2\g} f)(x) \propto\int_{-\infty}^{x} \rmd x'(x-x')^{m-1-2\g}\p^m_{x'}f(x')$, $({}_\infty\bp^\g f)(x) \propto (-1)^m\int_{x}^{\infty} \rmd x'(x'-x)^{m-1-2\g}\p^m_{x'}f(x')$, with $m-1\leq\g<m$. In the special case $\g=1$, one obtains a second-order operator which, unique in the class \Eq{calcagni:K}, can be written as a quadratic form \cite{calcagni:frc3}:
\be\label{calcagni:Ka}
\cK_\a=\cD_\mu\cD^\mu=\frac{1}{\sqrt{v_\a(x)}}\de^{\mu\nu}\p_\mu\p_\nu\left[\sqrt{v_\a(x)}\,\,\cdot\,\right]\,,\qquad [\cK_\a]=2\,.
\ee
Other self-adjoint Laplacians of the form $\cK=\cD^2$ are, of course, $\cK=\sum_\mu\cK_{\g,\a}(x^\mu)\cK_{\g,\a}(x^\mu)$.


\subsection{Properties}\label{calcagni:pro}

\subsubsection{Fractional versus fractal}

Are fractional spaces fractals? To answer this question, we should first specify what we mean by fractal. There is no consensual rigorous definition of fractal in the mathematical community, except perhaps the one by Strichartz: ``I know one when I see one'' \cite{calcagni:St03b}. In general, one must resort to a descriptive approach listing several desirable properties, most of which, anyway, are violated by at least one counterexample in the literature.

A fractal should have, first of all, a fine structure, i.e., detail at every scale. This structure should also be irregular, meaning that ordinary differentiability is given up. 
Third, many fractals are self-similar, i.e., detail does not change with the scale. Finally, fractals may have noninteger dimension (we already have quoted several counterexamples to this properties).

Fractional spaces are continua, and no matter how much one zooms into them, one will always encounter detail. Therefore they have a fine structure, although of a very boring type. They are not differentiable in an ordinary sense, but only according to (some of) the rules of fractional calculus. Thus, the presence of a nontrivial measure makes fractional spaces endowed with a structure which, from the standpoint of ordinary calculus, we could call ``irregular.'' We can check quantitatively whether $\cE_\a^D$ is also self-similar and of noninteger dimension.

A similarity $\cS_i$ is a map such that the distance between two points is proportional to the distance between their image through $\cS_i$:
\be
\Delta[\cS_i(x),\cS_i(y)]= \la_i\Delta(x,y)\,,\qquad 0<\la_i<1\,,\qquad i=1,\dots,N\geq 2\,.
\ee
Given $N$ such maps, self-similar sets are defined as the union of their own image through $\cS_i$ \cite{calcagni:Hut81}:
\be\label{calcagni:sss}
\cF= \bigcup_{i=1}^N \cS_i(\cF)\,.
\ee
Many deterministic fractals are of this type. $\cE_\a^D$ is trivially self-similar in geometric coordinates, just like $\mathbb{R}^D$. For instance, the closed interval $[0,1]$ in $D=1$ is covered by two maps, ${\rm S}_1(q) := \la q$, ${\rm S}_2(q) := (1-\la)q+\la$, where $\la$ is arbitrary. Genuinely self-similar sets are specified by given similarity ratios (for instance, the Cantor set is defined by the above two maps with $\la=1/3$), so we cannot claim to have found a symmetry structure of fractional Euclidean space. In section \ref{calcagni:dsi} this triviality will be fixed. For the time being, we notice that $\cE_\a^D$ actually possesses a richer symmetry, since it is invariant under an {affinity} transformation:
\be\label{calcagni:potra0}
{q'}^\mu =\Lambda(q^\mu) := \Lambda^\mu_\nu q^\nu+{\rm a}^\mu\,,
\ee
where now coordinates are mixed and the constant term is a vector. Up to boundary prescriptions, it is not difficult to recognize the group of rotations and translations in Eq.\ \Eq{calcagni:potra0}.

\subsubsection{Hausdorff dimension}

\begin{figure}
\includegraphics[width=5.6cm]{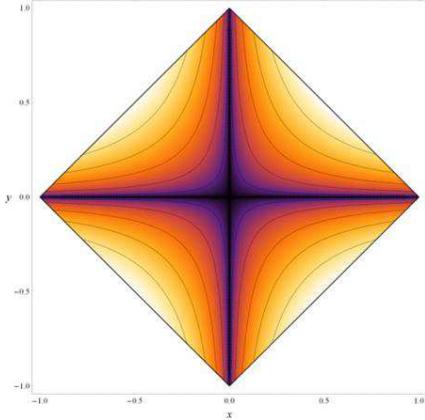}
\caption{Density plot of the area measure of a $1$-norm disk centered at the origin. The integration measure weight $v_\a$ is represented in light to dark shade, darkest shade being points where the weight diverges \cite{calcagni:frc1}.}
\label{calcagni:fig3}
\end{figure}

The Hausdorff dimension $\dh(\cE_\a^D)$ of Euclidean fractional space is nothing but the exponent in the scaling law \Eq{calcagni:scla}. An equivalent operative definition in the continuum is via the volume $\cV^{(D)}$ of a $D$-ball of radius $R$. Due to noninvariance under translations, the actual number depends on the location of the center of the $D$-ball (Fig.\ \ref{calcagni:fig3}). However, the radius dependence is universal and given by
\be
\cV^{(D)}(R)=\int_{D\textrm{-ball}}\rmd\vr_{\a}(x)\propto R^{D\a}\,,
\ee
leading to
\be 
\dh=D\a\,.
\ee
When $\a\neq n/D\in\mathbb{Q}$, the Hausdorff dimension is noninteger.

\subsubsection{Momentum transform}

Finally, let momentum space be endowed with a measure $v_{\a'}(k)$, with $\a'$ possibly different from $\a$. A unitary and invertible transformation between position and momentum space, expanded in a basis $\{\bE\}$ of eigenfunctions of the Laplacian \Eq{calcagni:Ka}, is \cite{calcagni:frc3}
\bs\label{calcagni:mot}\ba
&&\tilde f(k) := \int\rmd^Dx\,v_\a(x)\, f(x)\,\bE^*(k,x)\,,\qquad \bE(k,x)=\frac{\rme^{\rmi k\cdot x}}{\sqrt{v_\a(x) v_{\a'}(k)}}\,,\\
&& f(x) = \int\rmd^Dk\,v_{\a'}(k)\, \tilde f(k)\,\bE(k,x) \,,\qquad \cK_\a \bE(k,x)= -k^2 \bE(k,x)\,.
\ea\es
Compatibly with \Eq{calcagni:mot}, the fractional generalization of the delta distribution is not translation invariant:
\be\label{calcagni:deltas}
\de_\a(x,x') =\frac{\de(x-x')}{\sqrt{v_\a(x)v_\a(x')}}\,,\qquad \de_{\a'}(k,k') =\frac{\de(k-k')}{\sqrt{v_{\a'}(k)v_{\a'}(k')}}\,.
\ee
The momentum transform (or, more precisely, the infinite discrete class of transforms existing on $\cE_\a^D$) is an automorphism when $\a'=\a$.


\subsection{Spectral and walk dimension}

Given the above ingredients, a process diffusing a test particle from point $x'$ to point $x$ in fractional space is defined by the diffusion equation \cite{calcagni:frc1,calcagni:fra6}
\be\label{calcagni:die}
(\p_\s^\b-\ell^2\cK_\a)P(x,x',\s)=0\,, \qquad  P(x,x',0)=\de_\a(x,x')\,,\qquad 0<\b\leq 1\,,
\ee
where $\s\geq 0$ is a dimensionless diffusion parameter (a sort of artificial time or scale), $\ell$ is a length scale, the initial condition at $\s=0$ is associated with a pointwise (in the fractional sense of \Eq{calcagni:deltas}) probe, and we allowed for anomalous types of transport by taking the Caputo derivative $\p_\s^\b$ as the diffusion operator. Since there is only one scale, there is no hierarchy and the system is geometrically fixed.

One can attach a probabilistic interpretation to Eq.\ \Eq{calcagni:die} if $P\geq 0$ (which is the case \cite{calcagni:fra6}). Then, $P$ is the probability density distribution for a fractional Brownian motion on a fractal space. If $\b=1$ one simply has Brownian motion on a fractal. If $\b=1/2$, the process is an iterated Brownian motion (IBM) on a fractal. IBM's are models of diffusion of gases or liquids in cracks \cite{calcagni:crack}.

Letting the initial and final points coincide, one obtains the so-called return probabilty:
\be
\cP(\s):=\frac{1}{\int\rmd\vr_\a(x)}\int\rmd\vr_\a(x)\,P(x,x,\s)\,.
\ee
From this, one can show analytically that the {spectral dimension}
\be\label{calcagni:ds}
\ds := -2\frac{\rmd\ln \cP(\s)}{\rmd\ln\s}
\ee
is given by \cite{calcagni:fra6}
\be
\ds =\b\dh={\rm const}\,.
\ee
Notice that the formal definition of $\ds$ is independent on the details of the diffusion equation (including the order of the Laplacian).

The ratio of the Hausdorff and spectral dimensions yields the {walk dimension} $\dw= 2{\dh}/{\ds}={2}/{\b}$ \cite{calcagni:HBA}. Diffusion is classified as anomalous when $\ds\neq\dh$ (here, $\b\neq 1$). Superdiffusion ($\dw<2$, $\b>1$) does not correspond to fractals but to jump processes (e.g., \cite{calcagni:BGK}).

As anticipated, the spectral and Hausdorff dimension of space are constant and, in general, fractional. To get a more physical geometry, we must let the dimensions vary with the scale.


\section{Multifractional spaces}


\subsection{From fractional to multifractional}

Sets with scale-dependent geometry such as multifractals are described by self-similar measures. 
These are of the form $\vr(\cF)=\sum_{n=1}^N g_n\,\vr[\cS_n^{-1}(\cF)]$ \cite{calcagni:Hut81}.
One can think of a unit mass $\sum_n g_n=1$ distributed unevenly on subcopies of $\cF$, with probabilities $g_n$. For fractals with fixed dimension, the probability weights are all equal, $g_n=1/N$.

It is therefore immediate to define a multifractional action in the same manner:
\be\label{calcagni:ssa}
S=\sum_n g_n\int\rmd\vr_{\a_n}(x)\,\cL_{\a_n}\,,
\ee
where the Lagrangian density may itself depend on the fractional charges, for instance via the Laplacian $\cK_{\a_n}$. To check that the resulting space, multifractional Euclidean space $\cE_*^D$, has indeed a scale-dependent dimension, consider a simple model with only two terms (binomial measure). Integrals $I$ are made of two parts, which are combined via a dimensionful coefficient:
\be
I=I^{\a_1}+\ell_1^{D(\a_1-\a_2)}I^{\a_2}\,,\qquad [I]=-D\a_1\,,\qquad \tfrac12\leq\a_1<\a_2\leq 1\,,
\ee
where $\ell_1$ is a length scale. Calculating now the volume of a $D$-ball yields (the prefactors are the volumes of unit balls)
\be
\cV^{(D)}(R)= \ell_1^{D\a_1}\left[\Om_{D,\a_1} \left(\frac{R}{\ell_1}\right)^{D\a_1}+\Om_{D,\a_2}\left(\frac{R}{\ell_1}\right)^{D\a_2}\right]\,,
\ee
so that we can identify two geometric regimes:
\ba
{R\ll\ell_1}:&\qquad& {\cV^{(D)}\sim R^{D\a_1}}\,,\\
{R\gg\ell_1}:&\qquad& {\cV^{(D)}\sim \tilde R^{D\a_2},\qquad \tilde R=R \ell_1^{-1+\a_1/\a_2}}\,.
\ea
Here we defined the radius $\tilde R$ measured in ``macroscopic'' units. Thus, given a probing scale $\ell$, the Hausdorff dimension runs from $\dh(\ell\ll\ell_1)\sim D\a_1$ at small scales to $\dh(\ell\gg\ell_1)\sim D\a_2$ at large scales. In particular, setting $\a_2=1$ multifractional Euclidean space reduces, at large scales, to ordinary flat space in $D$ dimensions.


\subsection{Multiscale spectral dimension}

As in the fixed-dimension case, we analyze the fractional spacetime diffusion equation without metric corrections. The reason, as in Ho\v{r}ava--Lifshitz gravity, is that the anomalous character of empty flat spacetime is by itself sufficient to improve the UV behavior of the theory.

\subsubsection{Multiscale diffusion}

In correspondence with the self-similar action \Eq{calcagni:ssa}, we associate a multiscale diffusion equation where the Laplacian $\cK_\a$ is replaced by a sum over $\a$. As in known examples of multiscale complex systems, this sum is expected to be discrete. We call the coefficients $\zeta_n$, to possibly distinguish them from the $g_n$. Setting $\b=1$ for simplicity, we obtain
\be\label{calcagni:difef3}
\left(\p_\s-\sum_{n=1}^N\zeta_n\cK_{\a_n}\right)P(x,x',\s)=0\,,
\ee
with some initial condition $P(x,x',0)$. This system possesses $N-1$ characteristic scales $\ell_1<\ell_2<\dots<\ell_{N-1}$, and not $N$, since one of them must serve as probed scale. Typically,  the latter is the largest of the hierarchy, $\ell_N=\ell$, because one performs measurements via ``classical rods.'' Then, in \cite{calcagni:fra6} it was argued that $\zeta_N=1$ and
\be\label{calcagni:zeta}
\zeta_1(\ell)=\left(\frac{\ell_1}{\ell}\right)^{2}\,,\qquad
\zeta_n(\ell)=\left(\frac{\ell_n}{\ell-\ell_{n-1}}\right)^{2}\,.
\ee
In fact, $\zeta_n$ all have same scaling dimension and can be rendered dimensionless, as length ratios $\zeta_n=(l_{A,n}/l_{B,n})^q$. One can further choose $q=2$ to get a dimensionless multi-Laplacian $\sum_n(l_{A,n})^{2}\cK_{\a_n}$. Next, the $n$th term must dominate over the others at $\ell_{n-1}<\ell\ll\ell_n$, so $l_{A,n}=\ell_n$ and $l_{B,n}=\ell-\ell_{n-1}$. $l_{B,n}$ is chosen so that at $\ell\ll\ell_{n-1}$ the $(n-1)$th term takes over. Below $\ell_1$ there is no other scale, $\ell_0=0$. $\zeta_N\equiv1$ by definition. Therefore, dimensional flow is always measured starting from the lowest of two scales $\ell_{n-1}$ to the next $\ell_n$, and relatively to the latter, which sets a gauge for the rods.

An approximate solution to Eq.\ \Eq{calcagni:difef3} can be found by noting that one can reduce that to a single-scale equation with an effective fractional charge $\a_{\rm eff}$. For instance, in the $N=2$, $D=1$ case with $\a_2=1$,
\ba
(\p_x^2+\zeta_1\cK_{\a_1})P &=& (1+\zeta_1)\left[\p_x^2-\left(1-\frac{1+\zeta_1\a_1}{1+\zeta_1}\right)\frac{1}{x}\p_x+\frac{\zeta_1}{1+\zeta_1}\frac{(1-\a_1)(3-\a_1)}{4x^2}\right]P\nonumber\\
&=& \left[(1+\zeta_1)\cK_{\a_1(\ell)}+\frac{\zeta_1}{1+\zeta_1}\frac{(1-\a_1)^2}{4x^2}\right]P\,,\label{calcagni:useful}
\ea
where
\be\label{calcagni:prof1}
\a_1(\ell):=\frac{1+\zeta_1(\ell)\,\a_1}{1+\zeta_1(\ell)}\,,\qquad \zeta_1(\ell)=\left(\frac{\ell_1}{\ell}\right)^2\,.
\ee
The first term in \Eq{calcagni:useful} dominates for both small and large $\zeta_1$. There exists thus an effective fractional charge $\a_{\rm eff}\approx \a_1(\ell)$ throughout the dimensional flow. The error in the approximate solution is more pronounced at intermediate scales, where however one has a transient regime whose details are physically unimportant.

The general case with $\a_N=1$ is straightforward:
\be\label{calcagni:profN}
\a_{N-1}(\ell):=\frac{1+\sum_{n=1}^{N-1}\zeta_n(\ell)\,\a_n}{1+\sum_{n=1}^{N-1}\zeta_n(\ell)}\,,\qquad \zeta_n(\ell)=\left(\frac{\ell_n}{\ell-\ell_{n-1}}\right)^2\,.
\ee
This is the average $\langle\a\rangle$ of the coefficients $\a_n$ with respect to the weights $\zeta_n$. Consequently, the spectral dimension is
\be 
\ds\approx\ds(\ell)=\dh(\ell)=D\a_{N-1}(\ell)\,.
\ee

\subsubsection{Examples}

The simplest nontrivial example of dimensional flow features only one characteristic scale $\ell_1$:
\be
\left[\p_\s-\N_x^2-\left(\frac{\ell_1}{\ell}\right)^2\cK_{\a_1}\right]P(x,x',\s)=0\,,
\ee
where we set $\a_2=1$ in order to get $D$ dimensions in the infrared. Asymptotically, the spectral dimension reads
\be\label{calcagni:iruv}
\ds\sim\left\{ \begin{matrix} D\,,&\quad \ell\gg\ell_1\qquad \mbox{(IR)}\\
                              D\a_1\,,&\quad \ell\ll\ell_1\qquad \mbox{(UV)}\end{matrix}\right.\,.
\ee
In particular, \emph{if $\a_1=1/2$ is chosen as the minimum value of the range \Eq{calcagni:range} and $D=4$, $\ds\sim 2$ in the UV. Conversely, requiring a two-dimensional UV limit imposes $D=4$}. Here we are able to answer one of the questions posed in the introduction: Why 2 \emph{and} 4 dimensions? Demanding multifractional space to be normed at any scale and maximizing the excursion in the dimensional flow, one tightly links the deep UV to the far IR geometry.

Another single-scale system is the one described by the $D$-dimensional generalization of the Brownian-time telegraph process (e.g., \cite{calcagni:OB2}) $[\p_{\s}+(\ell_1/\ell)\,\p_{\s}^{1/2}-\ell_1^2\N_x^2]P=0$. Here the multiscale dependence is carried by the diffusion operator rather than the Laplacian. The resulting spectral dimension has qualitatively the same profile of the multifractional-space case, as discussed in \cite{calcagni:asf}.

\begin{figure}
\includegraphics[width=9.6cm]{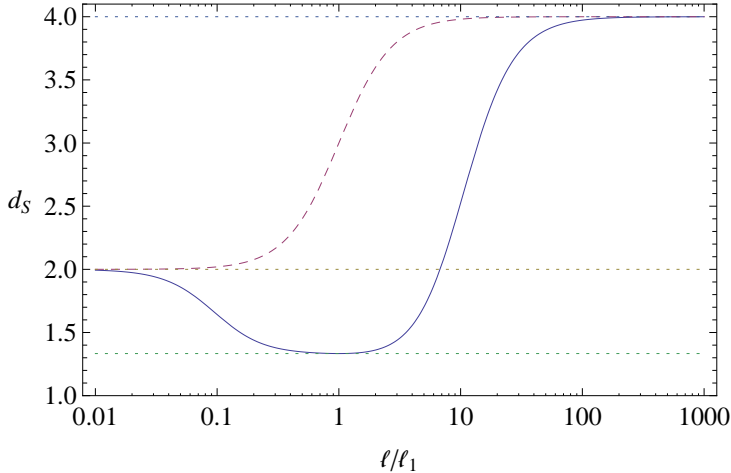}
\caption{\label{calcagni:fig4} The spectral dimension $\ds(\ell)$ in $D=4$ for a multifractional model and normal diffusion ($\ds=\dh$) with single scale (dashed curve) and two scales (solid curve) \cite{calcagni:fra6}.}
\end{figure}

Next, a two-scales system is governed by
\be
\left[\p_\s-\N_x^2-\zeta_1(\ell)\cK_{\a_1}-\zeta_2(\ell)\cK_{\a_2}\right]P(x,x',\s)=0\,,
\ee
leading to
\be\label{iruv3}
\ds\sim\left\{ \begin{matrix} D\,,                & \ell\gg\ell_2\gg\ell_1\qquad \mbox{(IR)}\\
                              D\a_2\,,& \qquad\quad~\ell_1\sim\ell\ll\ell_2\qquad \mbox{(intermediate)}\\
                              D\a_1\,,& \ell\ll\ell_1\ll\ell_2\qquad \mbox{(UV)}\end{matrix}\right. \,.
\ee
Figure \ref{calcagni:fig4} shows the single-scale and two-scale profiles with $\a_1=1/2$ and $\a_2=1/3$. The latter, effectively given by
\be 
\ds(\ell)=D\a_2(\ell)=D\frac{1+\frac12\zeta_1(\ell)+\frac13\zeta_2(\ell)}{1+\zeta_1(\ell)+\zeta_2(\ell)}\,,
\ee
closely resembles the one found in asymptotic safety in the presence of a cosmological constant and without matter \cite{calcagni:as,calcagni:asf}.


\subsection{Bounds on dimensional flow}\label{calcagni:bounds}

Although a systematic study of observational effects of multifractional theory is still lacking, one can make an educated guess about the size of the corrections expected in the UV and in the far infrared. Actually, several experimental bounds are known for toy models formulated in a dimensional regularization scheme where the dimension is $D-\e$ and the expansion parameter $\e$ is nonvanishing. Since geometric corrections in these theories are of the same order of magnitude as those in fractional spaces \cite{calcagni:frc1}, one can use bounds in dimensional regularization as a first approximation.

In the infrared, the anomalous magnetic moment $g-2$ of the muon gives an absolute upper bound, $|\e|\sim 10^3 |g_{\rm theor}-g_{\rm exp}|<10^{-8}$ at $\ell\sim 10^{-15}\,{\rm m}$ \cite{calcagni:Svo87}. At larger scales,  measurements of the Lamb shift for the hydrogen atom yield $|\e|<10^{-11}$ at scales $\ell\sim 10^{-11}\,{\rm m}$ \cite{calcagni:ScM}. Bounds at astrophysical and cosmological scales are weaker (from precession of Mercury, $|\e|<10^{-9}$ at $\ell\sim 10^{11}\,{\rm m}$ \cite{calcagni:ScM,calcagni:JY}; from pulsar measurements, $|\e|<10^{-9}$ at $\ell\sim 10^4\,{\rm ly}$ \cite{calcagni:JY}; from the cosmic microwave black-body spectrum, $|\e|<10^{-5}$ at $\ell\sim 14.4\,\mbox{Gpc}$ \cite{calcagni:CO}), but they constraint dimensional flow also in time.

In the deep UV, oscillations of neutral $B$ mesons and of the muon $g-2$ seem (this claim is isolated and further study is required) to suggest that any dimension between 2 and 5 is allowed by experiments at mass scales $M> 300\div 400~ {\rm GeV}$ \cite{calcagni:She09}, roughly corresponding to an upper bound $\ell_1<10^{-18}\,\mbox{m}$  for the multifractional model with monotonic dimensional flow. 


\section{Fields on fractional Minkowski spacetime}

Generalization of multifractional geometries to Lorentzian signature is straightforward. Fractional Minkowski spacetime $\cM_\a^D$ is defined by the set of data \Eq{calcagni:def}
with the embedding $\mathbb{R}^D$ replaced by $D$-dimensional Minkwoski spacetime $M^D$. The line element associated with $\cM_\a^D$ is $\rmd s^{2\a}=\eta_{\mu\nu} (\rmd x^\mu)^\a \otimes (\rmd x^\nu)^\a$ ($\eta_{\mu\nu}$ is the Minkowski metric) and is invariant under the fractional version of Poincar\'e transformations \cite{calcagni:frc2}
\be\label{calcagni:fpf}
{q'}^\mu = q({x'}^\mu)=\Lambda^\mu_\nu q^\nu(x)+{\rm a}^\mu\,.
\ee
Dimensions are calculated after Wick rotating the time direction, so again $\ds=\b\dh=\b D\a$.

A scalar field theory living on $\cM_\a^D$ is given by the Lagrangian density
\be\label{calcagni:scaa}
\cL_\a = \tfrac12\phi\cK_\a\phi-\sum_{n=1}^N\la_n\phi^n\,,
\ee
where we chose a polynomial potential. The scaling dimension of the field is such that
\be
[\phi]=\frac{D\a-2}{2} =0\quad \Leftrightarrow\quad \a=\a_1=\frac{2}{D}\,,
\ee
thus suggesting that in a deep UV regime where $\dh=2$ the theory is power-counting renormalizable. As usual, one classifies the operators $\cO$ in the action according to the scaling of their coupling $\la$:
\be
\cO=  \la\int \rmd\vr(x)\, \cO_d \sim \tilde \la \left(\frac{k}{E}\right)^{d-\dh}\,,\qquad  \la= \tilde \la E^{\dh-d}.
\ee
Relevant operators are important at low energies ($k/E\ll 1$) and the theory is said to be power-counting renormalizable if $[\la]\geq 0$ for all couplings. This happens when the dimension of all the operators in the action is $d\leq\dh$.

For the scalar theory \Eq{calcagni:scaa}, the condition $[\la_N]\geq 0$ implies
\bs\ba
&&N\leq \frac{2D\a}{D\a-2}\qquad {\rm if}\quad \a>\a_1\,,\\
&& N\leq+\infty ~~\quad\qquad {\rm if}\quad \a\leq\a_1\,.
\ea\es
Thus, at the UV critical point one has power-counting renormalizability. This result is confirmed by a computation of the superficial degree of divergence of Feynman diagrams, which in turn requires the scaling of the propagator (more generally, the Green's functions) \cite{calcagni:frc2,calcagni:frc6}.

Here we discuss neither the classical nor quantum dynamics of these models. We only notice that the number and form of relevant operators $\cO$ are constrained in the same way by RG and fractal-geometry arguments: the final total Lagrangians coincide. 

Quantum field theories on $\cM_\a^D$ or its multiscale generalization $\cM_*^D$ are not unitary. In fact, there is a loss of probability in the embedding bulk on general grounds \cite{calcagni:fra}. However, the loss of unitarity is under control at least for the free theory \cite{calcagni:frc5,calcagni:frc6}.

The same type of construction should apply also to gravity, for instance in the case of a Lagrangian density linear in the fractional generalization of the Ricci scalar, $\cL_\a\sim R^{(\a)}$. Power-counting renormalizability is then immediate, but of course proper renormalizability should be checked explicitly.


\section{Multifractional spaces and quantum geometry}\label{calcagni:mqg}

So far we have concentrated on the mathematical properties of multifractional spacetimes. Taken as fundamental, multifractional theory aims to be a candidate model of Nature borne out of the wish to remove UV divergences in perturbative gravity. However, by construction it can also serve as an effective description for other theories, since it is endowed with several geometric properties commonly found in quantum gravity scenarios. These properties are, in turn, mutuated from fractal geometry and multiscale complex systems. Given the generality of these tools, one should be able to describe multifractional spacetimes and quantum gravity at large with a similar language. We sketch their relation starting from the diffusion equation.

\begin{figure}
\centering
\includegraphics[width=9.6cm]{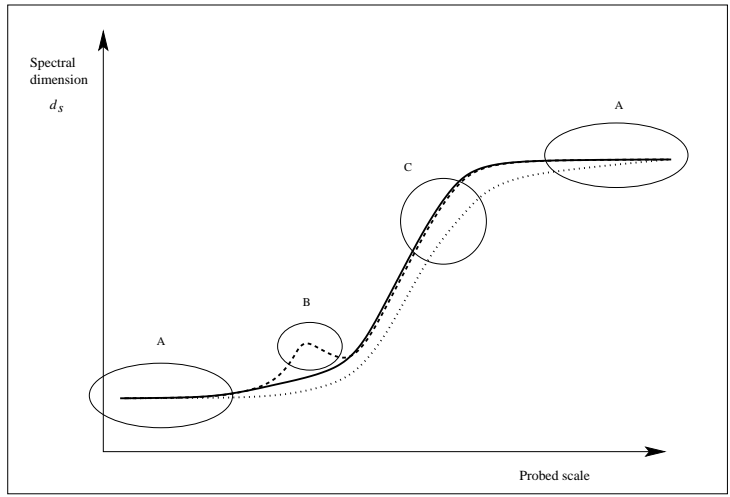}
\caption{\label{calcagni:fig5} Typical single-scale profile of the spectral dimension as a function of the probed length scale. (A) Asymptotic regimes where $\ds\sim {\rm const}$ (plateaux) and the values of $\ds$ therein are universal or almost universal. (B) Also intermediate plateaux, possibly reduced to local extrema, are robust within a given physical system, but different mathematical realizations of the same system cannot produce extra plateaux or transient features such as bumps, glitches, and so on. (C) Details of dimensional flow such as the monotonic slope of the profile between different regimes can change with the mathematical realization but they are physically unimportant \cite{calcagni:fra6}.}
\end{figure}

On a classical manifold, the diffusion equation is
\be\label{calcagni:cde}
\left(\p_\s-\N_x^2\right)P=0\,,\qquad {P(x,x',0)=\de(x-x')}\,,
\ee
corresponding to an ordinary Brownian motion and, in the absence of curvature, to a spectral dimension $\ds=D$. Classical gravity already modifies $\ds$, via the metric inside the covariant Laplacian. However, a quantum geometry can affect Eq.\ \Eq{calcagni:cde} even in regimes corresponding to zero curvature. There are three ways in which quantum effects can enter the diffusion equation:
\begin{itemize}
\item The diffusion operator can become fractional and multiscale, $\p_\s\to \sum_n\xi_n\p_\s^{\b_n}$. This may happen because quantum geometry modifies the scaling relation between momentum cutoff and diffusion parameter, as in asymptotic safety \cite{calcagni:asf}. Alternatively, anomalous scaling may be realized at the level of coordinates or metric, thus leading to a modified
\item Laplacian: $\N_x^2\to \sum_n\zeta_n\cK_{\g_n,\a_n}$. This is the case, again, of asymptotic safety \cite{calcagni:as,calcagni:asf} but also of Ho\v{r}ava--Lifshitz gravity, where the UV Laplacian is higher-order \cite{calcagni:hl}. The large class of simplicial gravities can also lead to a modified Laplacian in the continuum limit. The effective continuum Laplacian stemming from the limit of a generic simplicial pseudo-manifold can differ drastically from the naive continuum limit of the fundamental discrete Laplacian to Euclidean space. In fact, the effective continuum Laplacian can loose one or more of the properties of the discrete Laplacian, including locality and the differential order \cite{calcagni:COT1}.
\item The initial condition $\de(x-x')\to f(x,x')$ may also change, as in fractional spacetimes
or in effective models of ``quantum manifolds'' (first reference in \cite{calcagni:qs}).
\end{itemize}
Whether these modifications are explicit features of the given theory or are hidden by the technicalities of the latter is a model-dependent issue. Most of the theories share similar features in the profile of the spectral dimension (Fig.\ \ref{calcagni:fig5}). All have asymptotic plateaux, with same or very similar values. Intermediate regimes in one theory may disappear depending on certain details, such as the type of action and the presence of matter fields \cite{calcagni:asf}. On the other hand, transient monotonic phases can differ quantitatively (e.g., in the slope), but their exact form is not relevant because it relies on nonphysical details such as the employed regularization schemes.

Multifractional spacetimes and gravity theories based on the renormalization group approach (perturbative or nonperturbative) share many similarities in the way dimensional flow is realized. This happens because the RG flow can be described by tools of multifractal geometry, at least in a broad sense; therefore, since multifractional theory is founded upon the latter, one should expect to use it as a ``dual'' picture of RG-based theories. One of the crucial points is the identification of physical momentum. We briefly illustrate this in the following, leaving the details to the second paper in Ref.\  \cite{calcagni:asf}.

Consider first a $D$-dimensional gravitational field theory such that the effective Laplacian in the UV is higher order. For instance, in asymptotic safety higher-order Laplacians arise because the metric $g_{\mu\nu}$ scales anomalously with respect to the momentum cutoff $k$, $g_{\mu\nu}(k) = k^{-\delta} g_{\mu\nu}(k_0)$, where $\de$ is constant in each asymptotic regime (deep UV, far IR, and any intermediate plateau where $\ds\sim{\rm const}$) and $k_0$ is the IR reference scale. For simplicity we identify $k$ with the physical momentum, $k\equiv p_\textsc{qg}$.

If the theory is Lorentz invariant and the measure in position space is the usual one, in order to get a dual fractional picture ($\a$ is fixed for the time being) one should identify the coordinates $x_\textsc{qg}$ of this quantum gravity theory with the geometric coordinates $q$ of fractional theory, $x_\textsc{qg}^\mu=q^\mu$. Since $q\geq 0$, this identification is valid only in one orthant, which does not lead to a loss of generality for the purpose of this qualitative comparison. One could also define the coordinates $q(x)$ with an extra term sgn$(x)$ and extend the duality to the whole space. Setting $x_\textsc{qg}\leftrightarrow q$ is tantamount to ascribing anomalous scaling to a nontrivial measure weight acting as an effecting metric determinant, $v_\a(x)\leftrightarrow\sqrt{-g}$. Consistently, one obtains that $\a = 2/(2+\delta)$ and that the physical momentum $p_\textsc{qg}$ is conjugate to $q$, not $x$ (the standard coordinates of fractional spaces): $p_\textsc{qg} \sim x^{-\a}\propto q^{-1}$. As a consequence, the physical momentum $p_{\rm frac}$ of fractional theory is related to the momentum of the other quantum-gravity model by
\be 
(p_{\rm frac})^\a\sim p_\textsc{qg}.
\ee
Coordinates can be roughly assimilated to length scales measured by rods in a given theory: let $q\sim x_\textsc{qg}\sim L$ and $x\sim\ell$  define the ``length'' units for a given $\a$ (i.e., at a certain scale). Loosely speaking, in the generic quantum-gravity model and in fractional theory one has, respectively, ``$q$-rods'' measuring ``$q$-meters'' and ``$x$-rods'' measuring ``$x$-meters,'' mutually related by $L\sim\ell^\a$. ``Measurements by rods'' can also mean measurements of physical momenta.

In multifractional theory, geometric coordinates change with the scale (via $\a$), while $\{x\}$ and the physical momentum $p_{\rm frac}\sim x^{-1}$ are fixed. In particular, $x$-rods are fixed and they correspond to what we would call ``classical'' rods. In the RG-based theory, however, dimensional flow takes place by comparing the physics at any given scale $1/k=L$ with a classical length scale $1/k_0=\ell$. Therefore, the $q$-rod used in this theory must be $k$-adapted, yet $k_0$-dependent. To summarize, while in multifractional spaces the measure changes but not the rods/momenta, in the RG-based theory the measure is fixed but rods/momenta do change with the scale. On one hand, in multifractional models there is an ``integer'' classical observer measuring the fractal geometry at any scale with $x$-rods, which change intrinsically with the measure. On the other hand, in RG-based models the observer is ``fractal'' and measures the geometry with adapting $q$-rods. 

In the case of asymptotic safety, ordinary Lorentz invariance is preserved and the theory is physically different from multifractional models, where the action cannot be written as a standard $q$-dependent action with Lorentz-invariant Laplacian $\p_q^2$. Ho\v{r}ava--Lifshitz gravity is another example of RG-based (but perturbative) theory where the above picture holds, the only differences being that anomalous scaling is directly associated with coordinates (not the metric) and that it is not isotropic. Also in this case the Laplacian is none of the fractional Laplacians, even when the latter are considered in space-time anisotropic configurations (i.e., with different charges $a_0\neq \a_i=\a$, $i=1,\dots,D-1$). Therefore, we have not established a physical duality between multifractional spacetimes and RG-based theories. Rather, we have pointed out how RG physics can be recast in a language close to that of multifractal geometry and multiscale complex systems. Under this perspective, it is no wonder that Ho\v{r}ava--Lifshitz, asymptotic safety and multifractional theory share very similar profiles for dimensional flow.\footnote{This is true only in $D=4$. In three dimensions, Ho\v{r}ava--Lifshitz still predicts $\ds\sim 2$ in the UV, while asymptotic safety and isotropic multifractional theory coincide ($\ds\sim 3/2$). See \cite{calcagni:asf} for details.}


\section{Multifractional complex spacetimes}


\subsection{From real to complex order}

An important step beyond the multifractional setting presented so far is made after promoting the fractional charge $\a$ to a complex parameter \cite{calcagni:fra6}. Why to bother considering this extension of the theory? The reason is that real-order fractional integrals do not capture all the properties of genuine fractals. For instance, the return probability on deterministic fractals displays ripples, tiny oscilations due to the symmetry structure of these sets \cite{calcagni:osc}:
\be\label{calcagni:genKf}
\cP(\s)=\frac{1}{(4\pi \s)^{\frac{\ds}{2}}}F(\s)\,,\qquad  \text{$F$ periodic in $\ln\s$}\,.
\ee
Complex fractional integrals approximate integrals on fractals better than real-order calculus, inasmuch as the former approximate deterministic fractals, while the latter is better suited for random fractals. Complex-order calculus does include the logarithmic oscillations. Real-order fractional integrals are simply the average of complex integrals over a log-period \cite{calcagni:ff}.

\begin{figure}
\includegraphics[width=9.6cm]{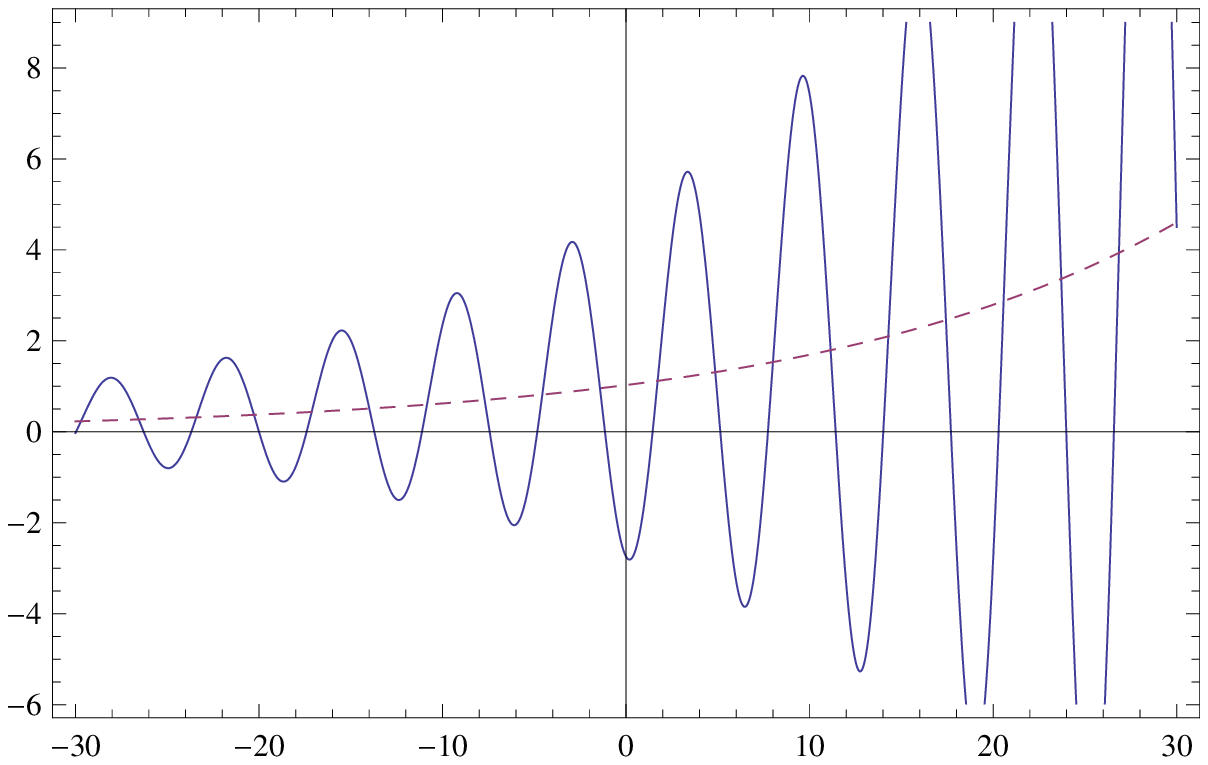}
\caption{\label{calcagni:fig6}The measure $\vr_{\a,\om}$, Eq.\ \Eq{calcagni:oscir}, for $\om=1$ and $\a=0.05$. The horizontal axis is $\ln|x/\ell_\infty|$. The dashed curve is the average $\vr_\a=\langle\vr_{\a,\om}\rangle$ \cite{calcagni:frc2}.}
\end{figure}

Complex fractional measures are obtained by the substitution $\a\to\a+\rmi\om$ in \Eq{calcagni:mea},
\be
\vr_\a(x)\to\vr_{\a,\om}(x)= c_+ |x|^{\a+\rmi\om}+c_- |x|^{\a-\rmi\om},\qquad \om\geq 0\,,
\ee
where $c_\pm$ are some coefficients. Summing over $\a$ and $\om$ and imposing the action to be real fixes $c_\pm$, so that one obtains 
\be
S=\int \rmd \vr(x)\,\cL\,,\qquad \rmd\vr(x) = \sum_\a g_\a\sum_\om\prod_\mu \rmd\vr_{\a,\om}(x^\mu)\,,
\ee
where
\be\label{calcagni:oscir}
\vr_{\a,\om}(x) = \frac{x^\a}{\Gamma(\a+1)}\left[1+A_{\a,\om}\cos\left(\om\ln\frac{|x|}{\ell_\infty}\right)+B_{\a,\om}\sin\left(\om\ln\frac{|x|}{\ell_\infty}\right)\right]
\ee
and $A_{\a,\om}$ and $B_{\a,\om}$ are real \cite{calcagni:frc2}. Here $\ell_\infty$ is a fundamental scale introduced to make the argument of the logarithms dimensionless. This form of the measure is also motivated by fractal-geometry arguments. Averaging over a log-period yields the zero mode of $\vr_{\a,\om}$, which is nothing but the real-order measure $\vr_\a$ (Fig.\ \ref{calcagni:fig6}).


\subsection{Discrete scale invariance}\label{calcagni:dsi}

The oscillatory part of $\vr_{\om,\a}$ is invariant under the transformation
\be
\ln\frac{|x|}{\ell_\infty}\,\to\, \ln\frac{|x|}{\ell_\infty}+\frac{2\pi n}{\om}\,,\qquad n=0,1,2,\dots\,,
\ee
implying the discrete symmetry
\be\label{calcagni:disi}
x\,\to\, \la_\om^n x\,,\qquad \la_\om:=\exp(2\pi/\om)\,,\qquad n=0,1,2,\dots\,.
\ee
Discrete scale invariance is typical of many chaotic systems ranging from earthquake models to financial crashes \cite{calcagni:Sor98}. As anticipated in Sec.\ \ref{calcagni:pro}, the trivial similarity symmetry of the real-order case is fixed once we move to a proposal in closer contact with fractal geometry.


\subsection{Scale hierarchy}

We can now identify several regimes from small to large scales. In the case of only one frequency $\om$ and for a monotonic dimensional flow (one characteristic scale $\ell_1$), we have:
\begin{itemize}
\item \emph{Ultramicroscopic regime ($\ell\sim\ell_\infty$).} Even if the formalism is in the continuum, log-oscillating measures render the geometry effectively discrete at scales close to $\ell_\infty$. Here, the measure is expanded about $|x|/\ell_\infty\sim 1$, getting $\rmd\vr(x)\sim \prod_\mu\rmd^D x/|x^\mu|$. This measure appears also in an apparently different context, $\kappa$-Minkowski noncommutative spacetimes. A close inspection of the relation between multifractional and noncommutative spacetimes \cite{calcagni:ACOS} makes this correspondence more precise, thus solving the problem of the commutative limit in $\kappa$-Minkowski (the cyclic-invariant measure $\prod_\mu\rmd^D x/|x^\mu|$ cannot reduce to the Lebesgue measure $\rmd^Dx$ because the Planck length is absent). From these results, one can naturally identify $\ell_\infty = \lp$.
\item \emph{Oscillatory transient regime ($\ell_\om=\lambda_\om\ell_\infty<\ell\ll\ell_1$).} At scales larger than $\ell_\infty$ but smaller than $\ell_1$, the geometry is described by Eq.\ \Eq{calcagni:oscir}. The system possesses a discrete scale invariance and ordinary geometric indicators such as dimensions and volumes are ambiguous unless averaged over a log-period $\ell_\om$. Without averaging, a determination of, say, the spectral dimension would yield widely different values by just a slight change in diffusion time. As a matter of fact, the correct definition of $\ds$, valid also for deterministic fractals, is the averaged version of Eq.\ \Eq{calcagni:ds}.
\item \emph{Multifractional regime ($\ell_\om\ll\ell\lesssim \ell_1$).} At mesoscopic scales (i.e., above the log-period but below $\ell_1$), the measure can be averaged:
\be
\vr_\a(x):= \langle\vr_{\a,\om}(x)\rangle\propto |x|^\a,\qquad \rmd\vr(x)\sim \sum_\a g_\a\rmd\vr_\a(x)\,.
\ee
Geometry effectively experiences a transition from a discrete to a continuum regime characterized, respectively, by the symmetries \Eq{calcagni:disi} and (at any given $\a$/scale) \Eq{calcagni:fpf}. There is a UV critical point at $\a=\a_1=2/D$, corresponding to $\dh=2$ and $\vr(x)\sim \vr_{1/2}(x)\propto |x|^{1/2}$. If $\a\geq\a_1$, then one must have $D=4$. Dimension flows from 2 to 4, and the IR dimension is fixed by the UV geometry.
\item \emph{Integer regime ($\ell\gg\ell_1$).} Eventually, at large-enough scales ordinary Poincar\'e-invariant field theory on Minkowski spacetime is recovered. The dimension of spacetime is $\dh=\ds=4-\e$ and Euclidean geometry in local inertial frames gets tiny corrections. We have already quoted some upper bounds on $\e$ in Sec.\ \ref{calcagni:bounds}. This regime may not necessarily be classical, i.e., one can enter it at scales still affected by quantum mechanics.
\end{itemize}


\section{Outlook}

Quantum gravity is a wide subject which can be explored in many different ways, not only in complex articulated frameworks such as string theory and loop quantum gravity, but also in other theories sharing some universal characteristics. Multifractional theory both is a model of quantum gravity in its own right and can serve as a framework to understand other proposals. Just like Ho\v{r}ava--Lifshitz gravity, it is a traditional perturbative field theory but built on an ``anomalous'' continuous spacetime. The development of the theory has just begun and is very much in progress. ``Euclidean'' and ``Minkowski'' classical geometries have been constructed; a sequence of scale and dimension hierarchies (including a discrete-to-continuum transition of geometry and, at larger scales, dimensional flow) are under analytic control; power-counting renormalizability has been shown; momentum space and invertible momentum transforms have been found; a detailed classification of diffusion and stochastic processes in quantum geometry was made possible; the relation with noncommutative spacetimes has been studied and $\kappa$-Minkowski geometry has been provided a novel embedding (as the asymptotic limit of a complex fractional geometry); quantum mechanics on fractional spacetimes has been also formulated. Among the works presently in progress, Noether currents and the quantum propagator of a scalar field are the next steps in the study of multifractional quantum field theory \cite{calcagni:frc6}; from this, one will be able to consider the issue of renormalization, the $\b$ functions, the level of quantum Lorentz violation, and particle-physics phenomenology.

One of the most pressing goals is to explicitly include gravity (and cosmology) in the picture, but we do not foresee any reason why it should not work \cite{calcagni:frc2}. (A preliminary analysis with gravity, based on a Lebesgue--Stieltjes formulation of the measure later abandoned for the technical reasons sketched in the introduction, is in \cite{calcagni:fra}.)  The purpose of the present paper, based on two lectures given at the Sixth International School on Field Theory and Gravitation in April 2012, was to elicit the reader's interest in the subject. We hope to have been successful and to report more results in the near future.






\bibliographystyle{aipproc}

\begin{thebibliography}{99}
\bibitem{calcagni:2e}   R.\ Gastmans, R.\ Kallosh, and C.\ Truffin, \doin{10.1016/0550-3213(78)90234-1}{Nucl.\ Phys.\ B}{133}{417--434}{1978}; S.\ Weinberg, ``Ultraviolet divergences in quantum gravity,'' in \emph{General Relativity, an Einstein Centenary Survey}, edited by S.~W.\ Hawking and W.\ Israel, Cambridge University Press, Cambridge, 1979, pp.\ 790--831; H.\ Kawai and M.\ Ninomiya, \doin{10.1016/0550-3213(90)90345-E}{Nucl.\ Phys.\ B}{336}{115--145}{1990}.

\bibitem{calcagni:df}  G.\ 't Hooft, ``Dimensional reduction in quantum gravity,'' in \emph{Salamfestschrift}, edited by A.\ Ali, J.\ Ellis, and S.\ Randjbar-Daemi, World Scientific, Singapore, 1993, pp.\ 284--296 [\oarX{gr-qc/9310026}]; S.\ Carlip, ``Spontaneous dimensional reduction in short-distance quantum gravity?,'' \doin{10.1063/1.3284402}{AIP Conf.\ Proc.}{1196}{72}{2009} [\arX{0909.3329}]; S.\ Carlip, ``The small scale structure of spacetime,'' in \emph{Foundations of Space and Time}, edited by G.\ Ellis, J.\ Murugan, and A.\ Weltman, Cambridge University Press, Cambridge, 2012 [\arX{1009.1136}].

\bibitem{calcagni:ncg} A.\ Connes, \tia{Noncommutative geometry and the standard model with neutrino mixing} \doij{10.1088/1126-6708/2006/11/081}{J.\ High Energy Phys.}{11}{081}{2006} [\oarX{hep-th/0608226}]; D.\ Benedetti, \tia{Fractal properties of quantum spacetime} \doin{10.1103/PhysRevLett.102.111303}{Phys.\ Rev.\ Lett.}{102}{111303}{2009} [\arX{0811.1396}]; E.\ Alesci and M.\ Arzano, \tia{Anomalous dimension in semiclassical gravity} \doin{10.1016/j.physletb.2011.12.026}{Phys.\ Lett.\ B}{707}{272--277}{2012} [\arX{1108.1507}].

\bibitem{calcagni:ACOS}  M.\ Arzano, G.\ Calcagni, D.\ Oriti, and M.\ Scalisi, \tia{Fractional
and noncommutative spacetimes} \doin{10.1103/PhysRevD.84.125002}{Phys.\ Rev.\ D}{84}{125002}{2011} [\arX{1107.5308}].

\bibitem{calcagni:sf} L.\ Modesto, \doin{10.1088/0264-9381/26/24/242002}{Class.\ Quantum Grav.}{26}{242002}{2009} [\arX{0812.2214}]; F.\ Caravelli and L.\ Modesto, \arX{0905.2170}; E.\ Magliaro, C.\ Perini, and L.\ Modesto, \arX{0911.0437}; G.\ Calcagni, D.\ Oriti, and J.\ Th\"urigen, work in progress.

\bibitem{calcagni:as}    O.\ Lauscher and M.\ Reuter, \doij{10.1088/1126-6708/2005/10/050}{J.\ High Energy Phys.}{10}{050}{2005} [\oarX{hep-th/0508202}]; M.\ Reuter and F.\ Saueressig, \doij{10.1007/JHEP12(2011)012}{J.\ High Energy Phys.}{12}{012}{2011} [\arX{1110.5224}]; M.\ Reuter and F.\ Saueressig, ``Asymptotic safety, fractals, and cosmology,'' in \emph{Quantum Gravity and Quantum Cosmology}, edited by G.\ Calcagni, L.\ Papantonopoulos, G.\ Siopsis, and N.\ Tsamis, Springer-Verlag, Berlin, to be published [\arX{1205.5431}].

\bibitem{calcagni:asf} G.\ Calcagni, A.\ Eichhorn, and F.\ Saueressig, work in progress; G.\ Calcagni, work in progress.

\bibitem{calcagni:cdt}   J.~Ambj{\o}rn, J.~Jurkiewicz, and R.~Loll, \doin{10.1103/PhysRevLett.95.171301}{Phys.\ Rev.\ Lett.}{95}{171301}{2005} [\oarX{hep-th/0505113}]; D.~Benedetti and J.~Henson, \doin{10.1103/PhysRevD.80.124036}{Phys.\ Rev.\ D}{80}{124036}{2009} [\arX{0911.0401}].

\bibitem{calcagni:hl}    P.\ Ho\v{r}ava, \doin{10.1103/PhysRevLett.102.161301}{Phys.\ Rev.\ Lett.}{102}{161301}{2009} [\arX{0902.3657}]; T.~P.\ Sotiriou, M.\ Visser, and S.\ Weinfurtner, \doin{10.1103/PhysRevLett.107.131303}{Phys.\ Rev.\ Lett.}{107}{131303}{2011} [\arX{1105.5646}].

\bibitem{calcagni:qs}    L.\ Modesto and P.\ Nicolini, \doin{10.1103/PhysRevD.81.104040}{Phys.\ Rev.\ D}{81}{104040}{2010} [\arX{0912.0220}];  E.\ Spallucci, A.\ Smailagic, and P.\ Nicolini, \doin{10.1103/PhysRevD.73.084004}{Phys.\ Rev.\ D}{73}{084004}{2006} [\oarX{hep-th/0604094}]; L.\ Modesto, \doin{10.1103/PhysRevD.86.044005}{Phys.\ Rev.\ D}{86}{044005}{2012} [\arX{1107.2403}]; T.\ Biswas, E.\ Gerwick, T.\ Koivisto, and A.\ Mazumdar, \doin{10.1103/PhysRevLett.108.031101}{Phys.\ Rev.\ Lett.}{108}{031101}{2012}
[\arX{1110.5249}]; L.\ Modesto, \doin{10.1103/PhysRevD.86.044005}{Phys.\ Rev.\ D}{86}{044005}{2012} [\arX{1107.2403}];  J.~R.\ Mureika, \doin{10.1016/j.physletb.2012.08.029}{Phys.\ Lett.\ B}{716}{171}{2012} [\arX{1204.3619}]; J.\ Mureika and P.\ Nicolini, \arX{1206.4696}.

\bibitem{calcagni:Bar83} J.~D.\ Barrow, \doin{10.1098/rsta.1983.0095}{Phil.\ Trans.\ Roy.\ Soc.\ Lond.\ A}{310}{337--346}{1983}. 

\bibitem{calcagni:mss}  G.~M.\ Zaslavsky, \tia{Chaos, fractional kinetics, and anomalous transport}
 \doin{10.1016/S0370-1573(02)00331-9}{Phys.\ Rept.}{371}{461--580}{2002}; D.\ Harte, {\it Multifractals: Theory and Applications}, Chapman \& Hall/CRC, Boca Raton, 2001.

\bibitem{calcagni:frc2}  G.\ Calcagni, \tia{Geometry and field theory in multi-fractional
spacetime}
 \doij{10.1007/JHEP01(2012)065}{J.\ High Energy Phys.}{01}{065}{2012} [\arX{1107.5041}].

\bibitem{calcagni:COT1}  G.\ Calcagni, D.\ Oriti, and J.\ Th\"urigen, \arX{1208.0354}.

\bibitem{calcagni:col} J.\ Collins, A.\ Perez, D.\ Sudarsky, L.\ Urrutia, and H.\ Vucetich, \doin{10.1103/PhysRevLett.93.191301}{Phys.\ Rev.\ Lett.}{93}{191301}{2004} [\oarX{gr-qc/0403053}]; J.\ Collins, A.\ Perez, and D.\ Sudarsky, \oarX{hep-th/0603002}.

\bibitem{calcagni:IRS}   R.\ Iengo, J.~G.\ Russo, and M.\ Serone, \doij{10.1088/1126-6708/2009/11/020}{J.\ High Energy Phys.}{11}{020}{2009} [\arX{0906.3477}].

\bibitem{calcagni:fra}  G.\ Calcagni, \tia{Fractal universe and quantum gravity}
\doin{10.1103/PhysRevLett.104.251301}{Phys.\ Rev.\ Lett.}{104}{251301}{2010} [\arX{0912.3142}]; G.\ Calcagni, \tia{Quantum field theory, gravity and cosmology in
a fractal universe} \doij{10.1007/JHEP03(2010)120}{J.\ High Energy Phys.}{03}{120}{2010} [\arX{1001.0571}]; G.\ Calcagni, \tia{Gravity on a multifractal}
\doin{10.1016/j.physletb.2011.01.063}{Phys.\ Lett.\ B}{697}{251--253}{2011}
[\arX{1012.1244}]. 

\bibitem{calcagni:frc3}  G.\ Calcagni and G.\ Nardelli, \tia{Momentum transforms and Laplacians in fractional spaces} \arX{1202.5383}.

\bibitem{calcagni:frc5}  G.~Calcagni, G.\ Nardelli, and M.\ Scalisi, \arX{1207.4473}.

\bibitem{calcagni:ff}   F.-Y.\ Ren, Z.-G.\ Yu, and F.\ Su, \doin{10.1016/0375-9601(96)00418-5}{Phys.\ Lett.\ A}{219}{59--68}{1996}; F.-Y.\ Ren, Z.-G.\ Yu, J.\ Zhou, A.\ Le M\'ehaut\'e, and R.~R.\ Nigmatullin, \doin{10.1016/S0378-4371(97)00353-1}{Physica A}{246}{419--429}{1997}; F.-Y.\ Ren, J.-R.\ Liang, X.-T.\ Wang, and W.-Y.\ Qiu, \doin{10.1016/S0960-0779(02)00211-4}{Chaos Solitons Fractals}{16}{107--117}{2003}; A.\ Le M\'ehaut\'e, R.~R.\ Nigmatullin, and L.\ Nivanen, {\it Fl\`eches du Temps et G\'eom\'etrie Fractale} (in French), Hermes, Paris, 1998; R.~R.\ Nigmatullin and A.\ Le M\'ehaut\'e, \doin{10.1016/j.jnoncrysol.2005.05.035}{J.\ Non-Cryst.\ Solids}{351}{2888--2899}{2005}.

\bibitem{calcagni:frc1}  G.\ Calcagni, {\it Adv.\ Theor.\ Math.\ Phys.} {\bf 16} (2012) (to be published) [\arX{1106.5787}].

\bibitem{calcagni:fra4}  G.\ Calcagni, \tia{Discrete to continuum transition in
multifractal spacetimes}
 \doin{10.1103/PhysRevD.84.061501}{Phys.\ Rev.\ D}{84}{061501(R)}{2011}
[\arX{1106.0295}].

\bibitem{calcagni:fra6}  G.\ Calcagni, \tia{Diffusion in quantum geometry} \doin{10.1103/PhysRevD.86.044021}{Phys.\ Rev.\ D}{86}{044021}{2012} [\arX{1204.2550}]; G.\ Calcagni, \arX{1205.5046}.

\bibitem{calcagni:HX}    H.-J.\ He and Z.-Z.\ Xianyu,
  \arX{1112.1028}.

\bibitem{calcagni:KST}   A.~A.\ Kilbas, H.~M.\ Srivastava, and J.~J.\ Trujillo, \emph{Theory and Applications of Fractional Differential Equations}, Elsevier, Amsterdam, 2006.

\bibitem{calcagni:BuPo} G.~L.\ Bullock, \tia{A geometric interpretation of the Riemann--Stieltjes integral} \ndoin{http://www.jstor.org/stable/2322483}{Am.\ Math.\ Mon.}{95}{448--455}{1988}; I.\ Podlubny, \tia{Geometric and physical interpretation of fractional integration and fractional differentiation} \ndoin{http://www.diogenes.bg/fcaa/}{Fract.\ Calc.\ Appl.\ Anal.}{5}{367--386}{2002} [\oarX{math.CA/0110241}].

\bibitem{calcagni:fdf}  K.\ Cottrill-Shepherd and M.\ Naber, \tia{Fractional differential forms} \doin{10.1063/1.1364688}{J.\ Math.\ Phys.}{42}{2203--2212}{2001} [\oarX{math-ph/0301013}]; V.~E.\ Tarasov, \tia{Fractional vector calculus and fractional Maxwell's equations} \doin{10.1016/j.aop.2008.04.005}{Annals Phys.}{323}{2756--2778}{2008} [\arX{0907.2363}].

\bibitem{calcagni:St03b} R.~S.\ Strichartz, \tia{Fractafolds based on the Sierpi\'nski gasket and their spectra} \doin{10.1090/S0002-9947-03-03171-4}{Trans.\ Am.\ Math.\ Soc.}{355}{4019--4043}{2003}.

\bibitem{calcagni:Hut81} J.~E.\ Hutchinson, \tia{Fractals and self similarity} \doin{10.1512/iumj.1981.30.30055}{Indiana Univ.\ Math.\ J.}{30}{713--747}{1981}.

\bibitem{calcagni:crack} K.\ Burdzy and D.\ Khoshnevisan, \doin{10.1214/aoap/1028903448}{Ann.\ Appl.\ Probab.}{8}{708--748}{1998}.

\bibitem{calcagni:HBA}   S.\ Havlin and D.\ Ben-Avraham, \tia{Diffusion in disordered media}
\doin{10.1080/00018738700101072}{Adv.\ Phys.}{36}{695--798}{1987}; D.\ ben-Avraham and S.\ Havlin, \textit{Diffusion and Reactions in Fractals and Disordered Systems}, Cambridge University Press, Cambridge, 2000.

\bibitem{calcagni:BGK} M.~T.\ Barlow, A.\ Grigor'yan, and T.\ Kumagai, \tia{Heat kernel upper bounds for jump processes and the first exit time} \doin{10.1515/CRELLE.2009.005}{J.\ Reine Angew.\ Math.}{626}{135--157}{2009}.

\bibitem{calcagni:OB2}   E.\ Orsingher and L.\ Beghin, \doin{10.1214/08-AOP401}{Ann.\ Probab.}{37}{206--249}{2009} [\arX{1102.4729}].

\bibitem{calcagni:Svo87} K.~Svozil, 
 \doin{10.1088/0305-4470/20/12/033}{J.\ Phys.\ A}{20}{3861--3875}{1987}.
 
\bibitem{calcagni:ScM}   A.~Sch\"afer and B.~M\"uller, 
 \doin{10.1088/0305-4470/19/18/034}{J.\ Phys.\ A}{19}{3891--3902}{1986}; B.~M\"uller and A.~Sch\"afer, 
  \doin{10.1103/PhysRevLett.56.1215}{Phys.\ Rev.\ Lett.}{56}{1215--1218}{1986}.
  
\bibitem{calcagni:JY}    C.~Jarlskog and F.\,J.~Yndur\'ain, 
 \doin{10.1209/0295-5075/1/2/002}{Europhys.\ Lett.}{1}{51--53}{1985}.
 
\bibitem{calcagni:CO}    F.~Caruso and V.~Oguri, 
 \doin{10.1088/0004-637X/694/1/151}{Astrophys.\ J.}{694}{151--153}{2009} [\arX{0806.2675}].
 
\bibitem{calcagni:She09} V.\,I.~Shevchenko, 
 \arX{0903.0565}.
 
\bibitem{calcagni:frc6}  G.~Calcagni and G.\ Nardelli, work in progress.

\bibitem{calcagni:osc}   M.\,L.~Lapidus and M.~van Frankenhuysen, {\it Fractal Geometry, Complex Dimensions and Zeta Functions}, Springer, New York, 2006; A.~Teplyaev,  \doin{10.1090/S0002-9947-07-04150-5}{Trans.\ Am.\ Math.\ Soc.}{359}{4339--4358}{2007}  [\oarX{math.SP/0505546}]; E.~Akkermans, G.\,V.~Dunne, and A.~Teplyaev,  \doin{10.1209/0295-5075/88/40007}{Europhys.\ Lett.}{88}{40007}{2009} [\arX{0903.3681}].

\bibitem{calcagni:Sor98} D.~Sornette, 
 \doin{10.1016/S0370-1573(97)00076-8}{Phys.\ Rept.}{297}{239--270}{1998} [\oarX{cond-mat/9707012}].

\end{thebibliography}

\end{document}